\documentclass[]{article}
\usepackage[margin=1in]{geometry}
\usepackage[hyphens]{url}
\usepackage{hyperref}
\usepackage[hyphenbreaks]{breakurl}
\usepackage{authblk}
\usepackage{amsmath,amssymb,amsfonts,amsthm}
\usepackage{algorithmic}
\usepackage{graphicx}
\usepackage{textcomp}
\usepackage[usenames,dvipsnames]{xcolor}
\usepackage{ragged2e}
\usepackage{hyperref}
\usepackage{multirow}
\usepackage{tabularx}
\usepackage{threeparttable}
\usepackage{booktabs}
\usepackage[normalem]{ulem}
\def\BibTeX{{\rm B\kern-.05em{\sc i\kern-.025em b}\kern-.08em
    T\kern-.1667em\lower.7ex\hbox{E}\kern-.125emX}}

\newtheorem{definition}{Definition}

\begin{document}

\title{Centering Policy and Practice: Research Gaps around Usable Differential Privacy\thanks{R.C. supported in part by NSF grant CNS-1942772 (CAREER), DARPA contract number W911NF-21-1-0371, and an Early Career Faculty Impact Fellowship from Columbia University. J.S. supported in part by the Columbia Data Science Institute and DARPA contract number W911NF-21-1-0371. Any opinions, findings, and conclusions or recommendations expressed in this material are those of the authors and do not necessarily reflect the views of the United States Government or DARPA.\\
The authors are grateful to Philip Leclerc, Priyanka Nanayakkara, Salil Vadhan, and Alex Wood for their comments on an earlier draft of this paper.}}

\author[1,2]{Rachel Cummings}
\author[2]{Jayshree Sarathy}
\affil[1]{Department of Industrial Engineering and Operations Research,
Columbia University}
\affil[2]{Data Science Institute, Columbia University}
\affil[ ]{\texttt{\{rac2239,js6514\}@columbia.edu}}

\date{}

\maketitle

\begin{abstract}
As a mathematically rigorous framework that has amassed a rich theoretical literature, differential privacy is considered by many experts to be the `gold standard' for privacy-preserving data analysis. Others argue that while differential privacy is a clean formulation in theory, it poses significant challenges in practice. Both perspectives are, in our view, valid and important. To bridge the gaps between differential privacy's promises and its real-world usability, researchers and practitioners must work together to advance policy and practice of this technology. In this paper, we outline pressing open questions towards building usable differential privacy and offer recommendations for the field, such as developing risk frameworks to align with user needs, tailoring communications for different stakeholders, modeling the impact of privacy-loss parameters, investing in effective user interfaces, and facilitating algorithmic and procedural audits of differential privacy systems.
\end{abstract}

\providecommand{\keywords}[1]{\textbf{\textit{Keywords ---}} #1}

\section{Introduction}\label{s.intro}

Statistical disclosure limitation has been an active area of research and practice for several decades~\cite{rubin1993statistical,matthews2011data}. With the introduction of \emph{differential privacy} (DP) in 2006~\cite{dwork2006calibrating}, however, the field has undergone an invigorating renewal. Differential privacy is a framework for adding noise to statistical releases in order to protect the privacy of individual data subjects. It provides a formal characterization of the tradeoffs between privacy and accuracy of statistical releases. Since its inception, differential privacy has been studied deeply in the theoretical computer science literature as well as in areas as diverse as statistics, demography, and law. DP has been deployed across government~\cite{abowd2018us} and industry~\cite{greenberg2016apple,erlingsson2014rappor,machanavajjhala2008privacy}, changing the way that organizations protect privacy of data subjects while enabling transparency and data sharing.

With these deployments have come significant challenges. The use of differential privacy has triggered confusion and contestation across nearly all of its deployments~\cite{boyd2022differential,tang2017privacy,greenberg2016apple,drechsler2021differential,agrawal2021exploring,steed2022policy,acquisti2023learning,santos2020differential,adeleye2023publishing,hod2024differentially}. 
Computer scientists have largely built consensus around the basic definitions and assumptions of this framework~\cite{altman2021hybrid}, but as DP spills out into other fields and engages a wide variety of stakeholders, its motivations, communications, design \& policy choices, and practical usage are questioned and negotiated anew by parties who understand data privacy from conflicting perspectives.  In our view, these frictions signal a need for researchers---both within and beyond computer science---to center policy and practice of differential privacy.
By policy, we are primarily concerned with \emph{organizational policy} rather than \emph{public policy}, although we believe that making progress on the former will help to inform the latter. We believe researchers should design guidelines around both technical and social factors of DP systems within organizational contexts. By practice, we are referring broadly to challenges related to the design, deployment, and use of DP systems.
This paper provides a roadmap of the open challenges and starting points for solutions to make DP significantly more usable in practice---and in some cases to develop alternatives that leverage the insights underlying DP's rigorous foundations yet are relaxed or tailored to meet the needs of stakeholders.

The goal of this paper is not just to outline the known challenges, but also to try to scope out the potential unknown hurdles to deploying DP. These unknowns are particularly important to address in order to regulate the use of DP and integrate the principles of DP into law and policy, as well as to prepare for the engineering challenges of large-scale DP deployments.

\section{Background and Overview}\label{s.background}

In this section, we provide a brief background on DP and an overview of the paper. We refer the curious reader to other surveys and textbooks for a more comprehensive overview of DP \cite{dwork2014algorithmic,wood2018differential,wood2020designing,near2021programming,cowan2024handson}. 

Differential privacy (DP), introduced by Dwork, McSherry, Nissim, and Smith in 2006~\cite{dwork2006calibrating}, is a mathematical definition of privacy that characterizes how much information a statistical analysis reveals about any one \emph{privacy unit} (typically, an individual or a small group of individuals) in the dataset. 
The standard definition is parametrized by two quantities---$\varepsilon$ and $\delta$---that denote the \emph{privacy loss} incurred by running a given set of analyses on the data.
A mechanism that limits privacy loss must introduce carefully calibrated noise to any computation over the data.

We provide the formal definition of DP below. Let $\mathcal{D}$ be a data universe and $\mathcal{D}^n$ be the space of datasets of size $n$.\footnote{This is called the \emph{bounded} DP definition because the size of the dataset is fixed and public. One could also consider the \emph{unbounded} definition for settings where the dataset size is itself sensitive~\cite{kifer2011no}.}
Two datasets $d, d' \in \mathcal{D}^n$ are neighboring, denoted $d \sim d'$, if they differ in a single record.\footnote{Another common formalization is that $d,d'$ are neighboring if $d'$ can be created from $d$ by adding or removing a single data record.}
Let $\mathcal{H}$ be a hyperparameter space and $\mathcal{Y}$ be an output space.	

\begin{definition}[Differential privacy \cite{dwork2006calibrating}]\label{def.dp}
A randomized mechanism $\mathcal{M}: \mathcal{D}^n \times \mathbb{R}_{\geq 0} \times [0,1] \times \mathcal{H} \rightarrow \mathcal{Y}$ is \emph{$(\varepsilon, \delta)$-differentially private} if for all datasets $d \sim d' \in \mathcal{D}^n$, privacy loss parameters $\varepsilon \geq 0$ and $\delta \in [0,1]$, hyperparameters $hp \in \mathcal{H}$, and events $E \subseteq \mathcal{Y}$,
\[
 \Pr[ \mathcal{M}(d, \varepsilon, \delta, hp) \in E] \leq e^{\varepsilon} \cdot \Pr[ \mathcal{M}(d', \varepsilon, \delta, hp) \in E] + \delta,
\]
where all probabilities are taken over the random coins of $\mathcal{M}$.
\end{definition} 

DP is an elegant theoretical formulation of privacy. It can account for current and future attacks, measure compositions of privacy loss over multiple data releases~\cite{dwork2014algorithmic}, and enable third-party scrutiny of the algorithm. This makes it possible for a data analyst to take into account the noise introduced when performing inference and estimating uncertainty through confidence intervals. 

Over the past two decades, disclosure methods that satisfy DP have been adopted by a variety of data-collection agencies and institutions, including Apple~\cite{greenberg2016apple}, Google~\cite{erlingsson2014rappor}, Uber~\cite{near2018differential}, the U.S. Census Bureau~\cite{machanavajjhala2008privacy,abowd2018us}, GovTech Singapore~\cite{290859}, the Wikimedia Foundation~\cite{adeleye2023publishing}, and the Israeli Ministry of Health~\cite{hod2024differentially}.\footnote{For a more comprehensive list of deployments, see~\cite{desfontaines2024list}.} In the last several years, we have also seen an emergence of open-source software libraries for DP data analysis, including DiffPrivLib~\cite{holohan2019diffprivlib}, OpenDP~\cite{gaboardi2020programming}, and Tumult Analytics~\cite{berghel2022tumult}. 

While these tools and deployments are promising, they have also made clear the difficulties of using DP in practice. We detail these challenges throughout the rest of the paper.
DP is no longer a new field---there has been a tremendous amount of work around DP algorithm design and statistical inference---but most of the literature has been focused on theoretical aspects. Only recently have scholars turned their attention to practical dimensions of using DP~\cite{agrawal2021exploring,garrido2022lessons,sarathy2023don,nanayakkara2022visualizing,cummings2021need,cummings2023challenges,xiong2020towards,bullek2017towards,near2021programming,ngong2023evaluating}, and research on these translational aspects are still in early stages.
In particular, there is a lack of consensus around policies and best practices of using DP across areas such as assessment of risk and incentives, communication, design, use, evaluation, and governance.
This paper aims to provide an overview of these understudied areas and develop a roadmap towards improving the practice of DP.

\subsection{Overview of findings}
We approach this paper as two computer scientists with over 15 years of experience, collectively, with the research and practice of DP. Our findings draw heavily on our professional experiences and expertise, as well as a review of the literature on making DP usable in practice.  Given our background, our findings are oriented towards the advancement of DP systems; however, we recognize that DP is not always the best approach, and we advocate for understanding the limits of DP and researching alternatives when relevant. 
Our overarching commitment is to privacy as a holistic, multifaceted concept. Our findings are also colored by our position as computer scientists; we regularly interact with data subjects, engineers, privacy officers, data curators, data analysts, executives, lawyers, and policymakers, but we are not experts in these areas. 

In this paper, we identify five important categories of concerns we must address to make DP usable in practice, now and in the future. These are: considerations of use; communication; design and policy; practice; and trust and governance. For each of these categories, we discuss challenges for researchers and practitioners, as well as some suggested directions for further inquiry and practical adoption. A summary of our findings are outlined in Table~\ref{tab:summary}. 
The last column of the table provides a brief assessment of how far along the community is in each of these categories. These categorizations are not meant to be exhaustive, but rather to serve as a starting point for future work in usable DP. 


\def\arraystretch{1.5}
\begin{table*}[htpb]
\hspace*{-1cm}
\centering
\begin{tabular}{|| p{2.4cm} | >{\raggedright\arraybackslash}p{4.5cm} p{7cm}|p{2.7cm} || }
 \hline 
 \textbf{Category} & \textbf{Challenges} & \textbf{Recommendations} & \textbf{State of the Field} 
 \\ [0.5ex] 
 \hline\hline
\multirow{4}{5em}{Considerations of Use (\S~\ref{s.whentouse})} & Assessing risk and privacy threats & Developing risk frameworks that align mathematical guarantees and user preferences & \multirow{4}{8em}[-6pt]{Requires shift in research agendas towards centering contextual analyses and stakeholder needs } \\ 
 & Understanding contextual needs & 
 Assessing informational norms and using DP attentively to context & \\
 & Aligning DP and institutional incentives & Positioning privacy as a positive good, while resisting privacy-washing & \\
 & Integrating DP with other privacy practices & Combining DP with cryptographic primitives and privacy assessments & \\
 \hline
 \multirow{3}{5em}{Communication (\S~\ref{s.comms})} & Tailoring communications for different parties & Understanding and addressing the needs of: data subjects, engineers, privacy officers, data curators, data analysts, lawyers, policymakers, and executives  & \multirow{3}{8em}[-4pt]{Promising initial research; future work should dive deeper into these questions} \\
 & Transparency traps &  Creating structures for making sense of technical transparency &\\
 & Engaging beyond $\epsilon$ & Communicating about \emph{all} the parameters and implementation choices involved in deploying DP 
 & \\
 \hline
 \multirow{3}{5em}{Design and Policy (\S~\ref{s.designandpolicy})} & Choosing definition and units & Developing guidance on translating risk assessments to these choices
 & \multirow{3}{8em}[-6pt]{Need to translate theoretical research into practical, actionable guidance} \\
 & Privacy-loss parameters & Modeling impact of parameters, testing explanations, creating databases, and translating to real-world risk & \\
 & Designing entire pipelines for DP & Considering impact of choices for sampling and editing data & \\
 \hline
 \multirow{3}{5em}{Practice (\S~\ref{s.practice})}  & Facilitating data processing and exploration & Creating best practices for dataset annotation and researching  strategies for exploration & \multirow{3}{8em}[-7pt]{Promising set of tools to build on, but requires more attention and investment } \\
 & Choosing metadata parameters & Building algorithmic and procedural toolkit for choosing parameters with limited domain knowledge & \\
 & Tools for using and evaluating data products & Investing in user interfaces, software libraries, and visualization tools & \\
 \hline
 \multirow{3}{5em}{Trust and Governance (\S~\ref{s.trustgovernance})} & Stakeholder engagement and oversight & Understanding how to meaningfully engage with stakeholders while lowering barriers for deployment & \multirow{3}{8em}[-4pt]{In very early stages; requires collaborative efforts and collective commitments} \\
 & Documenting decisions &  Building transparency and competitive pressure regarding deployment choices & \\
 & Auditing deployments & Enabling algorithmic and procedural audits & \\ [1ex] 
 \hline
\end{tabular}
 \\ [1ex]
  \caption{Summary of challenges, recommendations, and state of the field for using DP in practice}
  \label{tab:summary}
\end{table*}

\section{Considering when to use DP}\label{s.whentouse}

Before designing deployments of DP, organizations must first consider whether this technology is warranted and beneficial for their specific contexts and use-cases. This step is fraught with complexities around assessing risk, understanding practical privacy needs, considering institutional incentives, and integrating DP with other privacy tools.

\subsection{Assessing risk and privacy threats}

Building meaningful privacy solutions start with assessing privacy risks, but different communities understand the risks from releasing statistical information in vastly different ways. As researchers, we need to begin by understanding the types of risks and attacks that are important to stakeholders, and then analyze whether DP addresses these threats. This is a fundamental challenge because the framing of differential privacy corresponds to a particular understanding of privacy attacks as \emph{relative}, \emph{individual-focused}, and \emph{worst-case}, an understanding that may not be shared by all stakeholders, and one that represents a fairly recent shift in computer scientists' understanding of privacy attacks on statistical releases.

First, the DP guarantee considers \emph{relative privacy risk}, or the probability that an adversary can guess an individual's sensitive information based on a statistical release compared to their chance of doing so if that individual was not included in the dataset at all.
This baseline helps distinguish the informational impact of a given release from population-level inferences that could be made without access to the data, which is important for risk assessment both in computer science and law~\cite{altman2021hybrid}.
Stakeholders, however, may be more concerned with \emph{absolute privacy risk}, or the probability that the adversary can guess an individual's sensitive information regardless of the baseline success of the adversary when the individual is not included in the dataset. This discrepancy leads to disagreement about what constitutes effective privacy protection. For example, stakeholders have debated whether the use of aggregated census data to accurately infer an individual's race given their surname constitutes a privacy violation~\cite{kenny2021use}.
DP advocates would say no, as this inference could be made based on public knowledge of the statistical relationship between race and surnames.
Others might say that because of the potentially sensitive information that can be inferred about individuals, this is still a privacy violation---albeit one that is hard to protect against. Bridging these perspectives is critical for moving forward.

Second, the DP guarantee primarily considers the privacy of individuals or small groups. 
Some DP experts have stated that population-level statistical inferences are categorically \emph{not} privacy violations~\cite{bun2021statistical}.
However, this does not take into account that privacy is also relational and collective, as described by Viljoen as ``horizontal data relations"~\cite{viljoen2021relational} or by Barocas and Levy as ``privacy dependencies''~\cite{barocas2020privacy}. Extensions to the DP definition such as \emph{attribute privacy}~\cite{zhang2022attribute} have started to explore these directions; attribute privacy protects `global properties' of a dataset, such as the mortality rate of an entire hospital, as well as parameters of the distribution from which the dataset is drawn~\cite{zhang2022attribute}.
Other DP experts take a more moderate view, maintaining that while population inferences can be significant privacy risks, DP is still useful in that it allows us to \emph{separately} analyze individual and collective privacy risks and provide adequate protection to the former~\cite{wood2020designing}.

Third, DP provides guarantees of protection against a \emph{worst-case} adversary---one that can have complete knowledge of the dataset except for one individual's contribution. The DP guarantee ensures that even an adversary with this high level of knowledge about the dataset can gain only limited knowledge about the unknown individual. 
However, this guarantee is often too strong for the needs of stakeholders. Having such a strong guarantee comes at a cost; it is harder to satisfy while providing good accuracy. Stakeholders of DP deployments have argued that this theoretical worst-case notion of risk does not comport with their experience of real-world risks~\cite{ruggles2019differential}. Stakeholders may be more interested in the protections offered to each individual if we assume the adversary only has limited knowledge of the dataset. 

Advocates of DP point out that given the rise in availability of data sources and computational power, any assumptions about limits to the adversary's knowledge may not stand the test of time~\cite{wood2018differential}. 
But framing threats, risks, and motivation in these ways does not seem to connect with stakeholders who are used to thinking about privacy threats differently~\cite{ruggles2019differential,kenny2021use}.
In addition, the conservative guarantee of DP does not mean that it is a guarantee of the right `type'; using DP does not always provide insight into how well protected the dataset will be against the type of attacks that stakeholders \emph{are} worried about. These concerns will be specific to the application and context, but could include: absolute risk, group privacy, or even privacy concerns outside the scope of DP such as sharing data with third parties. Conversely, the protections that are within the scope of stakeholders' toolkits (e.g., heuristic approaches such as data suppression or rounding) are often too narrow and do not subsume the rigorous protections promised by DP~\cite{dwork2017exposed,altman2021hybrid}.
This is a major gap that needs to be addressed.

Given the fundamental differences in how DP frames risks relative to other understandings of privacy threats, a critical area for future research is around bridging these gaps. One avenue is to develop frameworks for assessing privacy threats and attacks that are quantifiable, rigorous, and attentive to future risks, but that are also meaningful to the particular context and goals of stakeholders. 
This is a challenging because of the space of parameters and choices specifying the risks that come from attacks will depend on the information and resources available to the adversary, the goals of the adversary, and other application-specific details \cite{CHSS24}. 
Collapsing these parameters into a single quantitative measure may not be informative, desirable, or even possible. Yet, the field would benefit from outlining the types of attacks that are important to stakeholders in each context and determine how DP addresses, or fails to addresses, these risks.

Additionally, research around formal privacy should give more attention to collective notions of privacy and the threats of inference beyond risks to individual privacy. As mentioned above, recent work has focused on applying a DP-like approach to different privacy units beyond individuals~\cite{zhang2022attribute}, and we believe such work is extremely important and fruitful.

\subsection{Understanding contextual needs}

Privacy is a broad, essentially contested concept~\cite{mulligan2016privacy}. It is only meaningful when considered in its social, legal, technical, and political context~\cite{nissenbaum2004privacy}. DP, on the other hand, is defined in a way that is agnostic to social context~\cite{benthall2022integrating}. It it is well-aligned with the aims of statistical disclosure limitation and enables concrete conversations around limiting privacy risks, but 
its algorithmic formalisms~\cite{green2020algorithmic} are not grounded in or attentive to its sociotechnical environment. 
While the abstract nature of DP facilitates theoretical analysis, it also creates a significant barrier to the deployment of DP in practice.

An open question is whether DP can be made more attentive to social and sociotechnical contexts. Regardless, it is clear that 
privacy harms need to be evaluated in a holistic manner and placed into broader contexts around legal standards and other forms of governance~\cite{nissim2017bridging,altman2021hybrid}. 
Using analytical approaches from contextual integrity~\cite{nissenbaum2004privacy} may be fruitful~\cite{benthall2022integrating}; for example, these methods can help elucidate why heuristic approaches to privacy, such as using methods of swapping or blank-and-impute, may not satisfy contextual obligations to protect against re-identification of data subjects in settings such as the U.S. census~\cite{abowd2018us}. In particular, the growing possibility for attacks against these heuristic approaches violate the contextual norms, as characterized by laws such as Title 13, around providing robust privacy protection to census respondents.

To be able to make these determinations, data owners need more guidance on assessing user expectations around privacy in a given context. Data curators and regulators need more detailed frameworks to understand what would need to change about existing data flows in order to conform to the contextual norms, and whether whether DP would allow data systems to conform to these norms or further undermine them~\cite{benthall2022integrating}. For example, suppose a company that collects user health data wants to improve its privacy practices. It currently shares aggregated  statistics about its users' health information with third parties such as insurance firms, and is considering using DP when sharing these statistics to better protect its users. To understand the impacts of DP on improving its privacy practices, the company should clarify the implications of its current data flows. What are the informational norms of this context and the current tensions with respect to (1) sharing aggregate statistics with third parties, (2) sharing aggregate statistics specifically with insurance companies, and/or (3) sharing aggregate statistics that can lead to individual harms for data subjects? And in which cases would using DP be beneficial?

In addition to understanding privacy and data sharing preferences, which are two aspects that have been explored in the literature~\cite{cummings2021need,nanayakkara2023chances,xiong2020towards,bullek2017towards}, we need better ways to understand other contextual needs, such as data utility, data accuracy, trust, explainability, and so on. In some cases, these other needs may indicate that DP is not appropriate for use. For example, a company may want to use DP to share data with third-party advertisers, but it may not be feasible for these advertisers to use noised estimates of market share or ad revenue. Similarly, if data for public use requires simple, explainable methods, DP may not be the right approach until improved explanations are developed. An important research direction is understand \emph{when} it is appropriate to use DP and what are the alternatives if not.

Finally, it is important to go beyond simply aligning DP with current norms or privacy expectations. Data practitioners should be equipped to make ethical judgments about what the norms \emph{should} be. Case studies that highlight the values embedded within modernizing privacy protections, such as Abdu et al.'s study of the U.S. census deployment of DP, are a valuable starting point~\cite{abdu2024algorithmic}. Normative thinking around DP includes identifying moral obligations to protect vulnerable stakeholders, and considering how DP would affect the ethical landscape of these norms~\cite{frischmann2014governing}. In our experience running working group sessions on responsible use of DP, we find that it can sometimes be hard for practitioners, especially technically focused ones, to engage in big-picture thinking around ethics and governance, but ethical guidelines such as the NIST Privacy Framework~\cite{NISTprivacyframework}, ASA Ethical Guidelines~\cite{AMSTATguidelines}, and UN Fundamental Principles of Official Statistics~\cite{UNSTATSguidelines} are valuable tools to help practitioners reason about the use of DP. Equally important are consent and oversight by ethical review boards that are well-equipped, and should be further trained, to make such judgments.

\subsection{Aligning DP and institutional incentives}

Contextual norms are one important avenue for evaluating privacy needs, but another approach is to consider the landscapes of institutional incentives for deploying DP. We identify three main challenges in this arena: perceived tensions, real incompatibilities, and practical barriers.

The first challenge is around the perceived tensions between privacy and corporate metrics. Privacy is often positioned as oppositional to data accuracy, innovation, and profitability. But as Julie Cohen argues, privacy is a positive good~\cite{cohen2012privacy}; it is critical for providing trust that leads to good data accuracy, space that enables innovation, and safety that protects good business. For example, officials at government agencies such as the U.S. Census Bureau understand that protecting privacy of data subjects \emph{now} is critical to maintaining data accuracy both \emph{now and later}. This is because without the promises and protections of privacy, communities---particularly those that are vulnerable or marginalized, such as non-citizens, victims of domestic violence, transgender youth, and tenants who could be evicted for not following occupancy mandates---are reluctant to provide their sensitive information to the Census Bureau. When the public lacks confidence in privacy protections, response rates go down, leading to differential undercounts and exacerbated errors for certain populations~\cite{mayer2002privacy}. In other words, there is little hope of accuracy without first having privacy. 

Positioning privacy as something that works in concert with other goals is a challenging endeavor. Research needs to explore how we can frame privacy as something that adds value, and something that is important to invest in, even if it comes at an initial cost. Privacy will almost always have some immediate benefits, and even more rewards in the future, such as safety, longevity, innovation, and utility~\cite{cohen2012privacy,mayer2002privacy}.

Of course, there \emph{are} some settings in which institutional logics are fundamentally in conflict with the goals of privacy. 
While using DP in these settings may protect data subjects from some privacy threats, it can also be used to hinder privacy goals more broadly~\cite{smart2022understanding}. For example, DP can be used to shut down stakeholders' privacy concerns (such as concerns about agency and consent) that cannot be translated into technical language, justify adjacent privacy harms such as extractive data collection, and reinforce centralized power~\cite{sarathy2022algorithmic}. Recent work has also highlighted the \emph{framing effects} of DP that can prioritize certain forms of risk management over others~\cite{seeman2022between}. For example, when using the local model of DP, the numerical measure of privacy loss cannot speak to the level of agency provided to data subjects in making data-sharing decisions and determining the privacy protections applied to their data. Consider a messaging application's use of local DP to collect statistics on what GIFs are most popular with users~\cite{tang2017privacy}. Data subjects' may be told that their sensitive data that captures keystrokes and GIF selections will be noised before being aggregated, but ultimately the application has control over how the protections are implemented and what parameters are used. Thus, the privacy-loss parameter leaves out important information about agency and control, which are crucial for understanding the privacy protections actually provided. 

With this in mind, we urge researchers to explore the following question: When do business incentives clash with the goals of privacy-enhancing mechanisms, and can DP really bridge the gaps? By identifying the tensions between corporate incentives and privacy tools, research can help us consider whether implementing DP would promote democratic engagement around data practices or, instead, be a tool for `privacy washing' or `privacy theater'~\cite{smart2022understanding}. 

Finally, there are many ways in which DP can be made more easily adoptable for engineering teams. DP primers and trainings that are catered not only to academics~\cite{wood2018differential,altman2021hybrid,heffetz2014privacy} but also engineers (such as~\cite{near2018differential}), are important for practical adoption. Equally important is the continued development of open source software libraries for DP data analysis, such as DiffPrivLib~\cite{holohan2019diffprivlib}, OpenDP~\cite{gaboardi2020programming}, and Tumult Analytics~\cite{berghel2022tumult}.  These and other libraries have already been influential in bringing together theory and practice of DP, and we need more such efforts towards building software that serves both as educational resources and trusted tools for development.  

Research should continue to focus on gaps for industry deployments and creating tools and algorithms to address real-world challenges. 
One of these challenges is around the many parameters that practitioners need to set in order to deploy DP successfully. As we discuss more in Section~\ref{s.designandpolicy}, this continues to be a hard problem.

\subsection{Integrating DP with other privacy practices}

The popular discourse around DP can sometimes seem to suggest that DP is a comprehensive privacy solution.
As most practitioners and experts know, however, DP is just one part of a suite of privacy and security tools. When considering the use of DP, organizations should also make sure to combine DP with best practices in security such as encrypted data storage, secure communication channels, and access control mechanisms.

Beyond this, there is promising research on  DP primitives that offer a wider set of protections or more favorable tradeoffs. For example, DP was first studied in terms of two models: the \emph{central model}, where noise is applied by a trusted data aggregator after they have collected sensitive information from data subjects, and the \emph{local model}, where data subjects add noise to their own individual bits of information before sending these to a potentially untrusted aggregator. While the local model requires much less trust, it does not allow for as much accuracy as in the central model.\footnote{For more details on the local vs. central model, see~\cite{desfontaines2021local}.} Recent work has put forth the \emph{shuffle model}~\cite{balle2019privacy,cheu2021differential}, where the data subjects' information is randomly permuted (but not noised) before reaching the aggregator. The shuffle model is innovative and valuable because it allows for accuracy that matches the central model with trust assumptions that are similar to the local model~\cite{erlingsson2019amplification}. 

The shuffle model was made possible by bringing tools and insights from cryptography to bear on DP algorithms. The success of this model---which uses just one of many cryptography primitives---suggests that researchers should be drawing more heavily on the intersection between DP and cryptography. As Wagh et al. write, such research is mutually beneficial to both communities, as cryptographic primitives can help bridge the utility gaps for DP, while DP relaxations of cryptographic primitives can yield implementations that are orders of magnitude faster than typical primitives~\cite{wagh2021dp}. Beyond shuffling, DP researchers can also look to multiparty computation~\cite{cramer2015secure,pettai2015combining}, zero-knowledge proofs~\cite{fiege1987zero}, and fully homomorphic encryption~\cite{gentry2009fully,li2022securing}; each of these points of intersection are vibrant areas of current research. We should continue to focus on how DP can interact with other privacy techniques to provide a more robust set of privacy protections.

\section{Communicating around DP}
\label{s.comms}

Once organizations have decided to use DP, they face the challenge of communicating about their deployment to relevant parties.  
In recent years, there has been a surge of research around explaining the protections of DP to end users~\cite{bullek2017towards,xiong2020towards,cummings2021need,smart2022understanding,franzen2022private,xiong2023exploring,karegar2022exploring,wen2023influence,nanayakkara2023chances,ashena2024casual}, but there is much more work needed in this arena, including: addressing the needs of different communities, avoiding transparency traps, and engaging with stakeholders on the many different deployment choices involved in DP.

\subsection{Tailoring communications for different stakeholders}
\label{ss.tailored-comms}

Communication around DP is not a monolithic endeavor. There are several parties involved in the implementation of DP---including data subjects, engineers, privacy officers, data curators, data analysts, executives, lawyers, and policymakers---and each require different types of tailored communications. Below, drawing on the suggestions from Cummings et al.~\cite{cummings2023challenges}, we describe communication challenges and areas for further research for each of these parties.

\begin{enumerate}

\item \textbf{Data subjects who decide to share personal data.} 

Data subjects are tasked with the decision of whether or not to contribute their sensitive data into an aggregated dataset or DP analysis. Making this decision in an informed manner requires understanding the protections afforded by DP in general, as well as the protections afforded by the specific privacy units, parameters, and model used by the system in question.
Recent work along these lines has aimed to evaluate data subjects' understanding of DP and data sharing preferences, given different types of explanations. For example, Cummings et al.~\cite{cummings2021need} find that the way DP is typically described is insufficient to help data subjects make informed decisions, and they recommend either training subjects to understand DP descriptions (shown by Xiong et al.~\cite{xiong2020towards} to be challenging) or to rely on approaches similar to privacy nutrition labels~\cite{kelley2009nutrition}. Nanayakkara et al.~\cite{nanayakkara2023chances} look more deeply into how data subjects understand privacy loss parameters (i.e., epsilon), offering suggestions for explaining these parameters in ways that maximize data subjects' risk comprehension. 
Further research on explanations---such as extending these findings to new data contexts and data uses---is critical for designing effective communications for data subjects.

Beyond understanding privacy risks and protections, data subjects also need to understand the upsides of contributing their data. Communications around DP should focus not just on the harms but also on the social and relational benefits of data sharing, as discussed by Viljoen~\cite{viljoen2021relational}. These themes have been touched on in prior work (e.g.,~\cite{bullek2017towards}), but explanations of the individual and social utility from DP releases have been under-explored in the literature thus far.

Finally, data subjects typically have the least agency over the DP system itself, compared to the many other stakeholders involved. It is important to find modes of communication---as well as concrete actions---that afford more power to data subjects over the use of their data in DP systems.

\item \textbf{Engineers who build and manage DP systems.} 

Engineers and managers (including technical and product managers) are at the core of DP deployments. They are involved at a hands-on level in ensuring that the system is reliable and that the privacy guarantees are correct. They are in charge of managing privacy composition over multiple statistical releases and verifying
privacy-accuracy tradeoffs. Communicating with this set of users should be technically oriented, but must go beyond creating academic materials. 

Current efforts to develop open-source DP libraries and repositories can help engineers understand what goes on under the hood in order to adapt methodologies for their own deployments. Even so, it can be challenging to get up to speed with a whole new paradigm under tight deadlines. In addition, recent user studies of open-source libraries for DP have demonstrated challenges around balancing ease-of-use and safe handling of data~\cite{ngong2023evaluating,song2024opensource}. Therefore, libraries should focus both on creating sound, vetted algorithms (e.g.~\cite{gaboardi2016psi}) and providing higher-level APIs that mimic the workflow of common data analysis libraries, such as numpy or scikitlearn
~\cite{bressert2012scipy} (e.g.~\cite{holohan2019diffprivlib}).

In addition, these APIs should incorporate visualization tools. Recent work on such tools~\cite{nanayakkara2022visualizing,budiu2022overlook} is promising  and calls attention to the many avenues for future work, such as providing visualizations for different types of queries, and enabling engineers to visually understand the impact on error of choosing metadata parameters, such as ranges and categories of data values.

\item \textbf{Privacy officers who are responsible for safe releases.}

Privacy officers (including privacy teams and disclosure review boards) are well-versed in traditional privacy protections, such as removing PII~\cite{mccallister2010guide}, swapping outliers~\cite{dalenius1982data}, and suppressing small counts~\cite{zayatz2007disclosure}. They have likely heard of DP but may not be sure it is right for the organization's context. Therefore, communicating with privacy officers should begin with explaining the particular situations in which DP is useful, as well as the situations that are outside of the scope of DP~\cite{wood2018differential,wood2020designing}. Communication should also emphasize the new risks that have been identified under traditional (or existing) privacy protections~\cite{dwork2017exposed,narayanan2008robust}. Finally, the research community should create trainings geared specifically towards privacy officers, empowering them to make decisions about parameter selection that are right for their organization, and giving them the tools to communicate with the other stakeholders in this list.

\item \textbf{Data curators who manage access to sensitive data.}
\label{ss.data-curators}

Data curators, who may be data scientists, researchers, statisticians, or administrators, have either collected or overseen the collection of the sensitive data, and are responsible for sharing it in an appropriate manner. Due to their detailed knowledge of the data domain and collection processes, they are likely in charge of setting global privacy-loss parameters and allocating privacy-loss budget to interested parties. Recent work studying data curators' decision-making around DP, however, has shown that making these decisions can be incredibly challenging~\cite{sarathy2022algorithmic}. The research community should create trainings and establish best practices to help curators make these important decisions.

Data curators are also involved with other data management practices, such as data annotation, archiving, and application/review/approval processes~\cite{mclure2014data}.
All of these practices may need to be modified when integrated with DP.
Dataset annotation, in particular, has been shown to be very important for releasing DP statistics~\cite{sarathy2023don}. Creating detailed annotations and codebooks is an important duty of the data curator in order to increase usability of the released statistics. For example, the curator should provide data bounds for numerical variables and a list of categories for categorical variables. They should also provide pointers to similar data that are publicly available. Sarathy et al.~\cite{sarathy2023don} also recommend that data curators set aside a portion of the privacy-loss budget for default releases of summary statistics, such as CDFs and histograms of key variables. However, more research is needed in order to provide guidelines for data curators on how to make such default releases and what portion of the privacy budget should be set aside for this.

The data curator must also fully contend with the downstream uses of data---not just what statistics are required, but \emph{how} the data are used. Data production is not a neutral endeavor~\cite{miceli2022data}; each method will entrench certain norms. For example, privacy protections used in the U.S. census prior to 2020---such as \emph{swapping}~\cite{zayatz2007disclosure,dalenius1982data}---added noise to the data but did not allow the bureau to reveal how the noise was added, thus entrenching the norm of treating census data as fact without accounting for error due to privacy protections~\cite{boyd2022differential}. When DP was introduced and data users were asked to take privacy error into account, this disrupted many downstream uses of census data.
DP may disrupt other norms around data use beyond accounting for error, such as accessing data via a query interface rather than direct exploration, and having a limited number of attempts to analyze the data due to a finite privacy loss budget versus unlimited data exploration~\cite{sarathy2023don}.
The data curator is responsible for explaining the limits of the data to stakeholders, especially when insights are shared using DP, so research needs to focus on best practices for communication along these lines.

\item \textbf{Data analysts who use DP releases for business, research and policymaking.}

Data analysts range from academic economists to grade school students. Communications with data analysts must be attuned to and matched with the level and type of technical detail that are used in their daily work. This suggests a strategy of layered communications, including: visual explanations of DP concepts to be used a high level, code snippets that demonstrate how to work with DP releases, and mathematical explanations that can be referenced if a data analyst needs to dig into the system processes. 

Communicating error is an important piece of these communications. It is natural for data users to want to overlook the error inherent in any data product, especially when they are tasked with using it to communicate with decision-makers.  DP, however, forces data users to take error into account. A DP release (whether in the form of query-release outputs or tabular synthetic data) is not a deterministic function of the collected data. This is sometimes readily apparent---if the data contain implausible values such as negative counts---but it can also take less obvious forms. Either way, data users must take into account the noise from DP releases when using them for decision-making. It is therefore important to find ways to communicate error to data analysts, and to allow them to remain cognizant about this error in their downstream uses of the data. 

\item \textbf{Executives who must decide on the use of DP and defend its impacts.}

The goals and incentives of executives may differ from those of engineers, privacy officers, and data curators. Executives are typically more focused on improving organizational performance along
metrics such as revenue, market share, or consumer satisfaction. They are unlikely to adopt DP purely for safety or ethical concerns.
Therefore, privacy needs to be communicated as a benefit rather than a moral imperative or a tradeoff with key performance metrics of the organization. This might look like highlighting demand for privacy protections from users, possibilities for higher quality and quantity of data collection, protection against real attacks from adversaries, avoidance of liability or subpoenas, compliance with privacy laws, and competitive advantages.

\item \textbf{Lawyers who must understand how DP systems fit with existing laws and regulations.} 

Lawyers are responsible for understanding whether DP is necessary and sufficient to meet the requirements of existing privacy and data processing laws. 
They often must determine whether a move to DP is justified by privacy regulations, and whether a new DP system is compliant with privacy laws. There are a few challenges of making these determinations. First, lawyers must translate between the mathematical language of DP and the legal formalizations of privacy, which can be misaligned in several ways, such as having continuous versus binary notions of risk; being agnostic to context versus specifically pertaining to data type, domain, and geography; and envisioning worst-case attacks versus attending to reasonable adversaries. There has been an innovative line of research in the past several years that translates between legal and mathematical notions of privacy~\cite{altman2021hybrid,nissim2017bridging,cohen2020towards}; this needs to be expanded to keep up with the growing adoption of DP across sectors and industries. 

\item \textbf{Policymakers who develop new regulations around data and privacy.}

Policymakers need improved communication around the tensions between computational understandings of risk and current privacy regulations. With guidance from both lawyers and computer scientists, they must understand the underspecifications in current privacy laws and how these can be redeveloped to remain in-step with technological advances and new understandings of privacy risks. Legal guidance will be especially important here to develop regulations that are contextually specific, yet also broad enough to remain relevant in the years ahead. In particular, experts in law and computer science must work together to design regulations and policy guidance based on concepts that are both legally and mathematically sound~\cite{altman2021hybrid,nissim2021foundations}. 

\end{enumerate}

\subsection{Avoiding transparency traps}\label{ss.transparency}

Organizations often equate effective communication with transparency. This is especially true with DP. Unlike other privacy techniques like swapping that rely on 'security through obscurity,' DP enables organizations to provide transparency about implementation details. 
Therefore, with DP it is not only \emph{possible}, but also \emph{encouraged}, to be fully transparent about the technical details of a privacy-preserving system to stakeholders such as data subjects, data users, and auditors.
However, this is not as straightforward as it seems. Prior deployments of DP have demonstrated that technical transparency requires supplementary tools and structures in order to be effective~\cite{boyd2022differential}.

Transparency can pose difficulties for communication in several ways. 
First, transparency around DP often requires justifying the \emph{need} for DP. As we discuss in Section~\ref{s.whentouse}, this is a complicated conversation, as stakeholders and organizations can have very different perspectives about what constitutes a privacy risk, what is a meaningful privacy attack, and whether DP is an appropriate intervention to address these threats.

Second, transparency around DP can hinder consensus across stakeholders. For example, highlighting the privacy-utility tradeoff, and the zero-sum nature\footnote{This can be a challenge with privacy design in general, see, e.g.,~\cite{cavoukian2009privacy}.} of choosing privacy-loss parameters and allocating them across data users, puts communication with stakeholders on a contentious playing field. 
In particular, relevant parties are more aware of how increasing accuracy of a particular statistic for one set of data subjects can decrease accuracy for another set of data subjects, since the privacy-loss budget is a finite resource. 
This happened with the communications around the U.S. census, when considering how much privacy-loss budget would be given to tribal communities, for example, where the push towards transparency created unexpected challenges for getting meaningful engagement from stakeholders about their privacy and utility needs~\cite{hawes2020implementing}. 
Second, communication via technical transparency requires shared background and perspectives around statistical concepts. For example, assurances to stakeholders that the scale of noise being added by DP is smaller than the scale of noise due to other sources (such as sampling, editing, and privacy mechanisms used prior to DP) requires first ensuring that stakeholders are aware of 
the noise inherent in all data processes~\cite{boyd2022differential}.

As discussed in Section~\ref{ss.tailored-comms} regarding data curators, the failure to account for stakeholders' comfort towards noise in data was one of the key communication hurdles during the U.S. census deployment of DP. When being transparent about the technicalities of DP, it is important to consider what was previously \emph{not} made transparent (e.g., the noise that was always included in census data due to sampling or data processing errors), the norms and data practices that were built upon faulty assumptions (e.g., policies around political redistricting in the U.S. that assume census data can be treated as accurate down to the individual respondent), and how transparency around DP upends these practices.

Third, transparency can create undue focus on some parameters over others. While there are many different design choices around DP, some of these choices (such as the choice of privacy-loss parameters) are more visible to stakeholders and easier to communicate about than others (such as sampling method or the choice of hyperparameters)~\cite{sarathy2023don}. We detail some of these design choices below in Section~\ref{s.designandpolicy}. It is important to consider which parameters are most important to get feedback on from stakeholders, while also not letting other impactful decisions get hidden in the process.

Transparency without adequate accompaniments to make sense of information, such as trusted expertise~\cite{abdu2024algorithmic}, can actively damage trust and communications. Therefore, it is important for researchers to understand what tools, practices, relationships, and supports are necessary for enabling stakeholders to benefit from transparency rather than become overwhelmed by inaccessible details. These are not new questions; they are studied heavily in the fields of science communication, design, journalism, and technology studies. Collaborations with researchers in these areas -- particularly researchers who study techniques for communicating uncertainty -- will greatly improve our own understandings of how to communicate effectively about DP.

\subsection{Engaging beyond epsilon}

While technical transparency comes with its challenges, it is still essential for organizations to be open and precise about the details of their DP systems. As Dwork, Mulligan and Kohli write, ``unless pertinent information is made available, assessment and comparison of DP systems is infeasible.'' They recommend creating an \emph{Epsilon Registry} for documenting and comparing technical implementations of different deployments. The goals of such a registry are to support shared learning across organizations, enable oversight, and exert pressure on organizations to provide meaningful privacy protections rather than using DP simply as privacy theater. 
Registries of this sort have been used by the Census Bureau~\cite{garfinkel2020status} and are called for by the broader DP community~\cite{desfontaines2024list}.

Despite the name of the registry, the authors note that it is critical to look beyond just privacy-loss parameters such as epsilon. As we have also discussed earlier in Section~\ref{ss.transparency}, there are several parameters that are critical for characterizing privacy and utility that should also be made available to the public, such as sampling methods, composition of algorithms, DP model (e.g., local, central, shuffle), algorithms used, and algorithm-specific hyperparameters. Communications regarding all of these implementation details---and the justifications for the choices---are critical for advancing the state of privacy protections via DP.

\section{Design and policy choices}\label{s.designandpolicy}

There are many design and policy decisions involved in using DP in practice. Below, we discuss these choices in more detail, and offer suggestions for easier decision-making.

\subsection{Choosing the right definition and unit of privacy}

The initial steps of a DP deployment include choosing a definition and unit of privacy. There are various definitions of DP, each with their own parameters that capture different aspects of privacy risk, including: pure $\varepsilon$-DP, approximate $(\varepsilon, \delta)$-DP, $(\mu,\tau)$-concentrated DP~\cite{dwork2016concentrated}, $(\rho)$-zero-concentrated DP~\cite{bun2016concentrated}, $(\varepsilon, \alpha)$-Renyi DP~\cite{mironov2017renyi}, $\mu$-Gaussian DP~\cite{dong2022gaussian}, and more. There are also various units of privacy, which refers to what DP will protect and what is considered a neighboring database. For example, the typical unit of privacy is for an individual, which means that a function run on a dataset \emph{with} this individual's data and a neighboring dataset \emph{without} this individual's data will result in similar (distributions of) outputs. In other words, DP will limit the influence of that individual's data on the final output. 

When an individual contributes multiple data points---such as a user interacting repeatedly with a platform---there is also a distinction between obscuring one single action of a user (\emph{event-level privacy}) or their entire history of actions (\emph{user-level privacy}). Other privacy units include small groups of individuals (\emph{group privacy}), data attributes when each individual's data have multiple features (\emph{attribute privacy},~\cite{zhang2022attribute}).

Making these decisions is foundational to the nature of the privacy protections. However, it is challenging to understand what is the right definition and unit to choose for a given deployment. System designers require more detailed guidance on how to translate privacy risk considerations into privacy definitions and units, how to understand the conversions from one set of definitions to another, and how to consider tradeoffs in making these decisions.

\subsection{Selecting and allocating privacy-loss parameters}

One of the main decisions the data curator must make is around privacy loss parameters. The leading privacy loss parameter in Definition \ref{def.dp} is epsilon ($\varepsilon$). Privacy theorists previously recommended that epsilon be set to be a small value between 0.01 and 1, although some practical deployments have necessitated using a much higher value~\cite{tang2017privacy,hawes2020implementing}, which corresponds to weaker privacy protections. 

The DP literature often approaches the problem of selecting the privacy-loss parameter in terms of optimizing the privacy-accuracy tradeoff~\cite{dwork2014algorithmic,geng2020tight}, which is a tricky problem itself. However, the job at hand is even more complex. To make this decision, data owners must integrate many factors including the needs of different stakeholders, prioritization of vulnerable stakeholders and scenarios, moral and ethical standpoints around privacy protections, risk calculations, changing needs over time as data systems scale, and so on. In addition, they must understand how these different factors translate to quantitative parameters. There is a considerable lack of guidance available to practitioners at the moment around making this choice. 

In addition to selecting the global privacy-loss budget, data curators must also allocate privacy-loss budget across different data analysts.
For example, research data repositories may want to allocate budget across different institutions with tiered levels of access, and companies may want to split the budget across internal analytics teams and third parties such as researchers or advertisers. Practitioners would benefit from principled methods and frameworks for conducting such allocations. 

Further research is required into how data owners can choose both the global privacy-loss budget and its allocation across stakeholders and use-cases. Such research can build upon existing work on modeling the impact of privacy-loss parameters using economic and social choice theory~\cite{abowd2019economic,hsu2014differential,kohli2018epsilon,dekel2023elasticity}, testing explanations of epsilon~\cite{mehner2021towards,nanayakkara2023chances}, creating databases that document parameters used across different deployments~\cite{dwork2019differential}, and translating parameters to real-world risk~\cite{pankova2022interpreting}. Beyond these approaches, it is also critical to bring in perspectives from social science and humanities to bear on this question. In particular, it will be important to understand how to integrate social and technical analyses when making decisions about a privacy-loss parameter that can impact a range of stakeholders in different ways.

\subsection{Designing entire pipelines for DP}

Much of the DP literature has focused on privacy loss starting at the point when data are shared (or sent by individuals to an aggregator), but it is increasingly important to understand how steps such as selecting the sample frame, sampling, weighting responses, imputing, and editing the collected data all affect the privacy guarantee. Researchers are starting to explore the design of entire data pipelines with DP in mind. 

Recent work has highlighted, for example, the ways in which sampling designs can both help and hinder the final privacy guarantee~\cite{balle2018privacy,bun2022controlling}. In particular, random sampling data points before a DP release can ``amplify'' the privacy guarantee, allowing one to gain more utility for the same privacy loss. Other types of sampling methods, such as proportional or cluster sampling, can have no effect on or even degrade the privacy guarantee. Those working with data would benefit from research that designs statistical methods to enhance, rather than degrade, the privacy protections offered by DP when considering the entire pipeline of creation and use.

DP pipelines should also extend well beyond the point of computation. DP outputs can take on a life of their own once they are released, and more research is needed to understand what constitutes a successful release \emph{after} the outputs are out in the wild. To ensure effective usage of DP statistics, we will likely need to build robust sociotechnical systems for citing, trusting, and explaining the outputs of DP mechanisms.

\section{Practice of DP}
\label{s.practice}
The next set of challenges we address concerns the practice of DP, which is only starting to be explored in the DP literature.
Practices of using DP include both what data users empirically do and what they \emph{should} do in order to create safe, usable releases.   

\subsection{Facilitating data processing and exploration}

Data processing and exploration are important steps in data analysis that are understudied in the DP literature.
Experts estimate that over 80\% of data analysis consists of data wrangling, cleaning, and editing~\cite{dasu2003exploratory}. However, these steps are difficult under DP constraints because it is unclear how they affect the privacy guarantee, and it is often challenging for data analysts to have enough context to perform these tasks when asked to do so via a privacy-preserving interface~\cite{sarathy2023don}.

Therefore, a critical area for future work is to build tools and strategies for processing and exploration under DP constraints. There are two main challenges to overcome: doing these processes without direct access the raw data, and conducting explorations with only a finite privacy loss budget.

One important strategy for mitigating these challenges is through detailed, thorough codebooks and data annotations. As discussed in Section~\ref{ss.data-curators},
data curators should provide information that enables data analysts to understand basic features of the data without spending additional privacy-loss budget on these steps. 
Further research is needed to understand what features are important to include in annotations, how annotations can be (semi)-automated, and best practices for creating and using codebooks during a DP release. 

A second important direction is regarding DP strategies for data cleaning and exploration. What parts of these processes can be systematically privatized, and what cannot? Many practitioners claim that these processes cannot be done with DP because they are not single algorithmic procedures, but rather a collection of tools, many of which involve looking at the data directly or searching for individual points (such as outlier detection) instead of global properties of the dataset. However, as datasets grow larger and contain millions of rows or features, practitioners are no longer able to just ``look at the data" to guide their analyses. As a starting point, some DP tools exist for data exploration tasks, such as privately computing marginal distributions of features, which can provide a picture of the dataset for analysts. Further research is needed to understand how to do other important data pre-processing tasks in a DP manner, such as checking for outliers that may represent miscoded data or dropping missing values.

\subsection{Choosing metadata parameters}
Datasets in the wild may have an infinite range or set of categories; this means the \emph{sensitivity} of analyses to the input data points could be unbounded, which can pose a problem for maintaining the privacy guarantee. Therefore, many DP algorithms require the inputs to be bounded, or for the curator/analyst to set a bound to which the input values can be clipped, in order to limit the privacy loss of the release~\cite{dwork2014algorithmic}. 

However, these bounds and categories must be data-independent; they cannot be taken straight from the dataset, but rather should come from knowledge about the data domain and its collection~\cite{near2021programming}. Selecting these parameters independently from the dataset is critical for maintaining the desired privacy guarantee. For example, consider a dataset containing the incomes of individuals in a company, and query for the mean income. Using the max income directly from the dataset for the upper bound of the `income' variable will directly reveal the sensitive income of this individual. 
However, these parameters also impact data utility. Ranges or categories that are too limited may introduce more bias into statistics computed, but parameters that are too broad may introduce more variance. Typically, the \emph{data curator} is responsible for choosing these parameters using their high knowledge of the data domain, but \emph{data analysts} are also accustomed to fine-tuning these parameters to achieve, e.g., their bias-variance goals.\footnote{For more details, see \url{https://programming-dp.com/ch5.html}.}

Across the board, practitioners of DP find that choosing metadata parameters in a way that preserves both privacy and utility is extremely challenging. Strategies do exist to mitigate this challenge, such as choosing these parameters based on publicly available data~\cite{seeman2020private,liu2021leveraging}, or spending a portion of the privacy loss budget on estimating these parameters privately (e.g., using DP binary search to estimate the range of a variable, as shown in~\cite{drechsler2022nonparametric})---although this may lead to other decision points, such as what portion of the privacy loss budget should be used. A goal of future work should be to create algorithmic and procedural strategies to guide setting of such parameters.

\subsection{Tools for using and evaluating data products with DP}

As mentioned in Section~\ref{s.comms}, transparency without adequate infrastructure can be counterproductive. We need pipelines for creating and evaluating DP data products (existing communities are starting to work on this, e.g.~\cite{berghel2022tumult}). This includes algorithmic research, such as providing DP confidence intervals and uncertainty measures that require minimal assumptions about the data (e.g.~\cite{ferrando2022parametric,wang2018differentially,drechsler2022nonparametric}), but it also includes  effective user interfaces, comprehensive software libraries~\cite{gaboardi2020programming,berghel2022tumult,holohan2019diffprivlib}, visualization tools~\cite{budiu2022overlook,nanayakkara2022visualizing}, and guidelines for citing and explaining the DP results to data users and policymakers.

\section{Trust and Governance of DP}
\label{s.trustgovernance}
Trust and governance of DP deployments requires bringing together all the components already discussed, from having reliable tools to regulating details of the deployment and considering legal and contextual alignments. This section will compile the different suggestions in prior sections toward creating more trusted---and trustworthy---DP deployments.

\subsection{Stakeholder engagement and oversight}
Organizations are increasingly interested in equity and participatory engagement~\cite{revesz2023}. However, it not always clear how to implement these ideals in practice, particularly when making decisions about technically complex choices such as those found within DP. Privacy co-design is challenging and depends heavily on context~\cite{friedman1996value,ackerman2001privacy,gurses2011engineering,cavoukian2009privacy}. Researchers have observed that---particularly with privacy, which is used as a catch-all concern for greater questions around data collection and governance---stakeholder concerns can either get lost in the design process or stall the deployment of the privacy technology completely~\cite{agrawal2021exploring}. We need more research and practice around what  it means to meaningfully engage with stakeholders and offer vulnerable parties both power and oversight, while still lowering barriers for important privacy projects to reach the stage of deployment.

\subsection{Documenting decisions, justification, and guidelines}
Documenting DP deployments is important for building trust.
Just as data curators should annotate datasets (e.g., using the framework of Gebru et al.~\cite{gebru2021datasheets}) prior to a DP release, organizations should also make public the justifications for their use of DP, alternative options that were considered and discarded, any relevant risks regarding privacy or data utility, level of protection against these risks that would be considered adequate for use case, whether DP addresses these risks, whether these risks and protections were discussed with stakeholders, and how each of these risks will be addressed in the deployment. These justifications should also bring together moral, legal, and technical guidelines around privacy in the organization's domain. The documentation should clearly outline all the decisions made in the design and implementation processes, compare these decisions with similar deployments, and provide explanations for the decisions (as recommended by Dwork et al.~\cite{dwork2019differential}). 

Finally, organizations should provide clear guidelines for use of the releases. What uses does the organization consider appropriate and aligned with the goals of the release? For example, the organization may encourage the use of DP statistics to learn about a group of consumers' behaviors, but not to target this group of consumers in harmful ways. Specifying appropriate use is also important for utility of DP releases, since some DP algorithms only provide strong accuracy guarantees on a pre-specified set of queries. 
While such guidelines will not always stop users from abusing the release, documenting the expectations of system designers is important for learning from the mistakes of each deployment and developing more robust guardrails over time.

\subsection{Auditing DP deployments}

As the use of DP becomes more mainstream, it will be increasingly important to have practices in place to audit these deployments. Such audits should start with the questions posed by Dwork, Kohli and Mulligan in their proposal of an \emph{epsilon registry}~\cite{dwork2009differential}, including: paths of privacy loss, granularity, epsilon per datum, burn rate, privacy loss allowed before retirement, variant of DP used, and justification for all of these implementation choices.
The responses to these questions should be compared with expert advice and across a range of deployments.

One challenge for trust and governance of DP comes from its inherent nature as a randomized algorithm, as well as the fact that multiple releases will continue to degrade privacy loss. When data users are only given one (or a few) releases from a single dataset, how can they verify that the algorithm was run correctly? Beyond institutional mechanisms such as regulations and audits, there is also room for technical methods that can provide trust in the DP release. This might look like zero-knowledge proofs of correctness, or mathematical proofs of privacy and utility. 

Finally, we need to answer the question: what recourse do data subjects (and/or data users) have if something goes wrong with a DP deployment? Answering this question requires dialogue between technical and legal stakeholders to understand what can and should be done in these cases.

\section{Conclusion}

DP is crossing the line from an emerging technology to an established one. With this shift comes new challenges and responsibilities for researchers and practitioners. Recent efforts have only scratched the surface. In this paper, we have outlined the next frontier of research around making DP usable. We center critical concerns around policy and practice, from assessing risks in a contextual manner to documenting and auditing deployments. We hope that the challenges and solutions identified in this paper can be a guide for researchers to focus their efforts, for policymakers to build frameworks to regulate DP, and for stakeholders to know where we must set our sights in order to align DP research with the problems it aims to solve. 
\clearpage

\bibliographystyle{abbrv}
\bibliography{bib}

\end{document}